\documentclass[twocolumn,showpacs,aip]{revtex4}
\usepackage{amssymb}
\usepackage{amsmath}
\usepackage{graphicx}
\usepackage{bm}
\usepackage{epstopdf}

\begin{document}




\title{Nonlinear plasma waves driven by  short ultrarelativistic electron bunches}

\author{Tianhong Wang$^1$, Vladimir  Khudik$^1$$^,$$^2$, Boris Breizman$^2$ and Gennady Shvets$^1$$^,$$^2$}
 \affiliation{$^1$School of Applied and Engineering Physics, Cornell University, Ithaca, New York 14850, USA.\\$^2$Department of Physics and Institute for Fusion Studies, The University of Texas at Austin, Austin, Texas 78712, USA.}


\date{\today}
\begin{abstract}
 We advance a theory of quasistatic approximation and investigate the excitation of nonlinear plasma waves by the driving  beam of ultrarelativistic electrons  using novel electrostatic-like particle-in-cell code. Assuming that the beam  occupies  an infinitesimally small volume, we find  the radius and length of the plasma bubble formed  in the wake of  the driver for varying values of the beam charge. The mechanism of the bubble formation is explained by developing  simple models of the bubble at large charges. Plasma electrons expelled by the driver charge excite  secondary plasma waves which complicate the plasma electron flow near the bubble boundary.
	%
	\end{abstract}

\maketitle

\section{Introduction}
%
Plasma-based accelerators are capable of producing much higher accelerating gradients than conventional radio frequency accelerating structures. In a plasma wakefield accelerator (PWA) an  electron bunch ~\cite{Chen_1985,Tajima_1979}
drives  a plasma wave propagating with phase velocity $v_{ph}\sim c$, where $c$ is the speed of light. This wave can create  a wakefield  accelerating gradient $E\sim\sqrt{n_0[cm^{-3}]}V/cm$, where $n_0$ is the density of plasma electrons. Recent experiments  have demonstrated rapid wakefield acceleration of electrons to very high energies~\cite{blumenfeld_energy_2007}
  in the strongly nonlinear  `plasma bubble'  regime~\cite{Rosenzweig_1991} with plasma electrons  fully expelled from the wake of the driving beam.

Particle-in-cell  (PIC) simulations~\cite{fonseca_osiris_2002,Pukhov_code, Nieter_2004}  is an indispensable tool for studying interaction of the driving beam with  plasma electrons, formation of the plasma bubble  and acceleration of the witness beam by the bubble fields. 
When the driver velocity $v_{dr}\rightarrow c$, 
the description of the plasma response and the evolution of electromagnetic fields 
 can be significantly simplified under the so called quasistatic approximation~\cite{Sprangle_1990,Mora_1997}. This approximation is a powerful approach laying the foundation for a number of quasistatic codes~\cite{Mora_1997,lotov_fine__2003,Huang_QUICK_2006} as well as simplified analytical models~\cite{Kostyukov_phenomenological_2004,Lu_theory_2006,Yi_Analytic_2013}. However the equation for the magnetic field involves time derivatives, and therefore is not completely quasistatic.  This deficiency complicates its numerical implementation   
and the overall physical picture.

The excitation of  nonlinear plasma waves by an  infinitesimally short driver   of  finite radius has been considered by Barov et al.~\cite{barov_energy_2004}, and the  momentum kick experienced by  plasma electrons from the passing driver   has been found analytically. Further simplification  of the driver has been considered in Ref.~\cite{stupakov_wake_2016}, where  the   beam radius is assumed to be the infinitesimally small. 
Such a `point-like' beam of ultrarelativistic electrons is characterized by only one quantity - its normalized charge 
\begin{eqnarray}
Q=\frac{k_p^3q}{4\pi e n_0},
\end{eqnarray}
where q is the beam charge,   $k_p\equiv\omega_p/c$, $\omega_p=(4\pi e^2n_0/m)^{1/2}$  is the electron plasma frequency, 
and $c$ is the speed of light. It turns out that in the  $Q\rightarrow 0$ limit the shape of the plasma bubble can be found analytically. 
 



Beyond the pointlike beam approximation, an analytical model  developed by Lu et al.~\cite{Lu_theory_2006}  predicts that  the plasma bubble in the wake of the  electron beam   resembles  a sphere at   $Q\gg 1$. This model involves an assumption that the bubble boundary can be approximated by the trajectory of a single plasma electron. It is however unclear whether  this  approximation is justified for a point-like driver. 

As we will see below, the pointlike driver with $Q\gg 1$ kicks plasma electrons at relatively small distances  from the driver $r\sim k_p^{-1}$. Yet, the size of the plasma bubble observed in simulations~\cite{Rosenzweig_2004}  is much larger than $k_p^{-1}$. 
This discrepancy and underlying physics mechanism of the bubble formation are explained in the present paper.
%
%
%

We  use the Vlasov equation for the electron distribution function 
under  quasistatic approximation  to describe the plasma response to the self-consistent electromagnetic field in the wake. We derive the quasistatic equations for  electromagnetic fields, which do not involve time derivatives in the source terms.  Thus, the problem is treated in the  similar way as    the   electrostatic one \cite{Birdsall_book}. 
%
%

We incorporate the quasistatic approximation in a novel  particle-in-cell (PIC) code. We investigate  nonlinear plasma waves in the wake of the pointlike driver and find the   radius and  length of the plasma bubble for arbitrary values of the driver charge $Q$. 
We also develop  phenomenological models for the plasma bubble in the $Q\gg 1$ limit and compare them with  models~\cite{Kostyukov_phenomenological_2004, Lu_theory_2006}.  

The rest of the paper is organized as follows. In Sec.~II we present the basic equations for the  quasi-static theory.
The main result of this section is the derivation of the Helmholtz-type equation for  magnetic field,
with a source term depending only on  radial positions and momenta of the plasma electrons.
 This  equation is used in Sec.~III to find the initial momentum kick from the driver. We then discuss the wakefield in a ballistic regime (Sec.~IV),   
and present results of the particle-in-cell simulations (Sec.~V).
 In Section~VI, we suggest  simple models of the plasma bubble behind a large charge driver. These models capture essential features observed in PIC simulations. Finally,  we summarize our findings  in Sec.~VII.    
%
\section{Equations of quasistatic approximation.}
This Section addresses  the dynamics of plasma electrons behind a pointlike ultrarelativistic electron beam propagating along a positive direction of the $z$-axis in a cold homogeneous plasma. We adopt a quasistatic approximation by assuming that the driver velocity is equal to the speed of light.

 In what following, we use dimensionless units normalizing time  to $\omega_p^{-1}$, length to $k_p^{-1}$, and velocities to  $c$. We also normalize the electron kinetic momentum ${\bf{p}}$ to $mc$, the fields ${\bf{E}}$ and ${\bf{B}}$ to $mc\omega_p/e$, the potentials $\phi$ and {\bf{A}} to $mc^2/e$, the plasma density to $n_0$, and the current density  ${\bf{j}}$  to $-en_0c$.
 In dimensionless units,  
 the current  and charge densities of the electron  beam   are given by
\begin{eqnarray}
&&j_{dr}=\rho_{dr}=
2(Q/r_{\perp})\delta(r_{\perp})\delta(t-z),\label{eq:driver}
\end{eqnarray}
where $r=\sqrt{x^2+y^2}$.

The 3-dimensional motion of plasma electrons in the wake of the driver depends on a single dimensionless parameter $Q$, and is adequately described by the  Vlasov kinetic equation for the electron  distribution function $f_e$  complemented by Maxwell's equations. We describe this motion under two assumptions: (I)   the  distribution function $f_e$ and electromagnetic fields depend on $t$ and $z$ only through a combination $\xi=t-z$, and (II) the  motion  is axially  symmetric.
 
First, we simplify the kinetic equation
\begin{eqnarray}
&&\frac{\partial {f}_e}{\partial t} +\frac{\partial H}{\partial {\bf{P}}}\cdot\frac{\partial f_e}{\partial {\bf{R}}}-\frac{\partial H}{\partial {\bf{R}}}\cdot\frac{\partial f_e}{\partial {\bf{P}}}=0,\label{eq:Vlasov_3D}
\end{eqnarray}
where  
$H=[1+({\bf{P}}+{\bf{A}})^2]^{1/2}-\phi$ is the Hamiltonian 
 ${\bf{P}}={\bf{p}}-{\bf{A}}$ is the canonical momentum, $\phi$ is the electric potential, and ${\bf{R}}=(x,y,z)$ is the electron radius vector.  The trajectory of an individual electron in  phase space $({\bf{R}},{\bf{P}})$ is determined by the equations of motion:
\begin{eqnarray}
&&\frac{d{\bf{P}}}{dt}=-\frac{\partial H}{\partial {\bf{R}}}, \quad \frac{d{\bf{R}}}{dt}=\frac{\partial H}{\partial {\bf{P}}}.
\label{eq:EoM_3D}
\end{eqnarray}

From the dependence of $H$ on $\xi=t-z$, we derive that $dH/dt={\partial H}/{\partial t}={\partial H}/{\partial \xi}$  and $dP_z/dt=-{\partial H}/{\partial z}={\partial H}/{\partial \xi}$. Hence, $H-P_z=const$. Due to cylindrical symmetry, the  Hamiltonian $H$ does not depend on the azimuthal angle $\theta$, so that $d{{P}}_{\theta}/dt={\partial H}/{\partial \theta}=0$. Hence, the azimuthal momentum ${{P}}_{\theta}=const$. Since all electrons start their motion from cold homogenous plasma where $H=1$ and ${\bf{P}}=0$, these integrals of motion take the form:
\begin{eqnarray}
&&H-P_z-1=0, \quad P_{\theta}=0.\label{eq:Conser_G}
\end{eqnarray}
In the cylindrical coordinates $(r,\theta,z)$, there are two  non-vanishing components of the electric field ($E_r$ and $E_z$), and one non-vanishing component of the magnetic field ($B_{\theta}$).  Using the   gauge $ {\bf{A}}_{\perp}=0$, we  express the magnetic field as $B_{\theta}=-\partial A_z/\partial r$ and  the transverse components of canonical momentum as ${{P}}_{\theta}= {{p}}_{\theta}=xp_y-yp_x$ and $P_r=p_r=(xp_x+yp_y)/r$.
%

%
 %

Using assumptions (I) and (II) and conservations laws (\ref{eq:Conser_G}),  the distribution function of electrons can be expressed in the form
\begin{eqnarray}
f_e(t,{\bf{R}},{\bf{P}})=rf_{*}(\xi,{{r}},{{P}}_{r})\delta(H -P_z-1)\delta(P_{\theta}),
\label{eq:substitute}
\end{eqnarray}
where $f_*$ represent a distribution function    of macroparticles performing one dimensional radial motion; the factor $r$ in front of $f_*$ is introduced to take into account the particle weight in the cylindrical coordinates, and the variable $\xi$ plays now a role of a new `time'.  
Substituting Eq.~(\ref{eq:substitute}) into (\ref{eq:Vlasov_3D}), we find that $f_*$ satisfies the one-dimensional Vlasov equation  (see Appendix A):
\begin{eqnarray}
&&\frac{\partial }{\partial\xi}{f_{*}} +\frac{\partial H_*}{\partial {{p}}_{r}}\frac{1}{r}\frac{\partial }{\partial {{r}}}(rf_{*})-\frac{\partial H_*}{\partial {{r}}}\frac{\partial }{\partial {{p}}_{r}}f_{*}=0,\label{eq:Vlasov_1D}
\end{eqnarray}
where 
\begin{eqnarray}
&& H_*(\xi, r,p_r)=\frac{1+{{p}}_{r}^2+(1+\psi)^2}{2(1+\psi)}-\psi-A_z\label{eq:H_1D}
\end{eqnarray}
is the  Hamiltonian for cylindrically symmetric motion~\cite{Mora_1997}. This Hamiltonian   depends on $\xi$ and  $r$ through the wakefield potential $\psi\equiv\phi-A_z$ and the vector potential $A_z$. The trajectory of an individual particle in the phase space $(r,p_r)$  is determined by equations of motion
\begin{eqnarray}
\frac{d}{d\xi}{p}_{r}=\frac{\gamma}{1+\psi}\frac{\partial}{\partial r}\psi -B_{\theta}, \quad
\frac{d}{d\xi}r=V(\xi,r,p_r), \label{eq:EoM_1r}
\end{eqnarray}
 where $\gamma=[{1+{{p}}_{r}^2+(1+\psi)^2}]/{2(1+\psi)}$ is the relativistic factor, and $V\equiv {p_{r}}/({1+\psi})$ is the particle `velocity' in $r$-direction.

Integration of  Eq.~(\ref{eq:Vlasov_1D}) over momentum $p_r$ gives the  continuity equation:
\begin{eqnarray}
&&\frac{\partial}{\partial \xi}n_{*}=
-\frac{1}{r}\frac{\partial }{\partial {{r}}}r {{n}}_{*}\langle V\rangle, 
\label{eq:continuity_1}
\end{eqnarray}
where    $n_{*}=\int d {{p}}_{r}f_{*}$ is the   density of macroparticles, and the brackets denote averaging over transverse momentum $\langle V\rangle={{n}}_{*}^{-1}\int d {{p}}_{r}f_{*}V$.
As seen from Eq.~(\ref{eq:continuity_1}) that the total number of  macroparticles is the same at each  $\xi=const$ slice. 

In addition to conservation  of macroparticles, the total energy of  macroparticles  is  conserved (see  Appendix B and Ref.~\cite{Mora_1997}). For the point-like driver, this additional conservation law has  the  form  
\begin{eqnarray}
2\pi\hspace{-1mm} \int_{0}^{\infty}\hspace{-2.5mm}rdr\bigg[n_*\langle\gamma-1\rangle+\frac{E_z^2+(\partial\psi/\partial r)^2}{2}
\bigg]=const.
\label{eq:conserv_energy1}
\end{eqnarray}

We also note that  the density and the current density of plasma electrons can be expressed through the distribution function  $f_*$ as:
 \begin{eqnarray}
n_e=\frac{n_*\langle \gamma \rangle}{1+\psi},\quad j_{r}=n_*\langle V\rangle, 
\quad j_z=\frac{n_*\langle p_z\rangle}{1+\psi},\label{eq:Jr}
\end{eqnarray}
 where $\langle p_z\rangle=\langle \gamma \rangle-\psi-1=[1+\langle p_r^2\rangle-(1+\psi)^2]/(2(1+\psi)$. Besides, $n_e-j_z=n_*$. 

We now show that the wakefield potential $\psi$  and the electric field  $E_z$  at a `time' $\xi$ are determined only by positions and momenta of macroparticles at the same time, that is, by  $f_*(\xi,r,p_r)$.
Using  $\xi$ as a time-like variable, Maxwell's equations in dimensionless variables take the following form:
\begin{eqnarray}
&&\nabla\times {\bf{E}}=-\frac{\partial}{\partial \xi}{\bf{B}},
\label{eq:ME_1}\\
&&\nabla\times {\bf{B}}=\frac{\partial}{\partial \xi}{\bf{E}}-{\bf{j}}.
\label{eq:ME_2}
\end{eqnarray}
We find from Eq.~(\ref{eq:ME_1}) that  $E_r=-\partial \psi/\partial r+B_{\theta}$ and $E_z=\partial \psi/\partial \xi$ in the axially symmetric problem. Substituting these expressions into Gauss's law $\nabla\cdot {\bf{E}}=-n_e+1$ and using  Eq.~(\ref{eq:ME_2}), we find the following equations:
\begin{eqnarray}
\frac{1}{r}\frac{\partial }{\partial r}r \frac{\partial }{\partial r}\psi=n_{*}-1, \quad 
\frac{\partial }{\partial r}E_z=-n_*\langle V\rangle.
\label{eq:Eq_for__psi}
\end{eqnarray}
 Their solution must satisfy the following boundary condition: $\psi\rightarrow 0$ and $E_z\rightarrow 0$ at $r\rightarrow \infty$.   Note that the source term $n_*-1$ in equation for $\psi$ is fully determined by macroparticles' positions while the source term $n_*\langle V\rangle$ in equation for $E_z$ requires knowledge of the macroparticles' positions and  momenta. 

Taking a curl of the both sides of Eq.~(\ref{eq:ME_2}) and  replacing $\nabla \times {\bf{E}}$ by $-\partial {\bf{B}}/\partial \xi$, 
 we find an equation for the magnetic field $\Delta_{\perp}{\bf{B}}=\nabla\times {\bf{j}}$. Its  angular component has the following form:
 \begin{eqnarray}
&&\frac{\partial }{\partial r}\frac{1}{r} \frac{\partial }{\partial r}r{{B}}_{\theta}=-\frac{\partial }{\partial r}j_z-\frac{\partial }{\partial \xi}j_r,
 \label{eq:Mag_axial}
\end{eqnarray}
where $j_z$ and $j_r$ are defined by Eq.~(\ref{eq:Jr}).  To obtain a closed form of the equation for magnetic field, we  establish a  relationship  between the time derivative of the transverse current  and the  electromagnetic fields. 
Such a relationship is well-known for cold weakly perturbed plasma~\cite{Keinigs_1986}: $\partial {\bf{j}}/\partial t=(\omega_p^2/4\pi){\bf{E}}$. To generalize it to   strongly perturbed relativistic plasma, we  multiply  Eq.~(\ref{eq:Vlasov_1D})
by the 'velocity' ${{V}}=\partial H_*(\xi,r,p_r)/\partial{p_r}$ and integrate it over momentum $p_r$. 
After straightforward calculations we obtain
\begin{eqnarray}
\frac{\partial}{\partial\xi}{{j}}_{r}=n_{*}\langle{{a}}\rangle-\frac{1}{r}\frac{\partial}{\partial r}rn_{*}\langle V^2\rangle\rangle,
\label{eq:Flux_axial}
\end{eqnarray}
where $a\equiv{d^2 r}/{d\xi^2}$ is the particle 'acceleration':
\begin{eqnarray}
&&{{a}}=-\frac{B_{\theta}}{1+\psi}+\tilde{a},\label{eq:accel_B_ax}\\
&&\tilde{a}=\frac{\gamma}{(1+\psi)^2}\frac{\partial\psi}{\partial r}-\frac{{{V}}}{1+\psi}\bigg(E_z+V
\frac{\partial \psi}{\partial {{r}}}\bigg).\label{eq:accel_psi_ax}
\label{eq:dpsi}
\end{eqnarray}
We herein define $\tilde{a}$ as a part of `acceleration' caused by the wakefield potential $\psi$ and $E_z$.
Finally, substituting Eqs.~(\ref{eq:Flux_axial}) - (\ref{eq:dpsi}) into Eq.~(\ref{eq:Mag_axial}) we find that the magnetic field satisfies the Helmholtz equation
\begin{eqnarray}
&&\frac{\partial}{\partial r}\frac{1}{r} \frac{\partial}{\partial r}rB_{\theta}=\frac{n_{*}}{1+\psi}B_{\theta}-S, 
\label{eq:B_final}
\end{eqnarray}
where $S$ is the `source' term 
\begin{eqnarray}
S=n_*\langle \tilde{a}\rangle+\frac{1}{r}\frac{\partial}{\partial r} rn_*\langle V^2\rangle+\frac{\partial}{\partial r}j_z.
\label{eq:S_r_}
\end{eqnarray}
Equation~(\ref{eq:B_final}) needs to be solved with the boundary conditions $B_{\theta}\rightarrow 0$ at $r\rightarrow \infty$ and $B_{\theta}\rightarrow 0$ at $r\rightarrow 0$.

Shielding of the magnetic field by  plasma is clearly seen from  Eq.~(\ref{eq:B_final}).  Indeed, the coefficient in front of the first term on the  right-hand side of this equation is proportional to $\delta_s^{-2}$, where $\delta_s$ is the skin depth of relativistic plasma. In dimensional variables,  
\begin{eqnarray}
&&k_p^2\frac{n_{*}}{1+\psi}=\frac{4\pi e^2n_e}{m\langle\gamma\rangle c^2} =\frac{1}{\delta_s^{2}}.
\label{eq:SKIN}
\end{eqnarray}
   
In quasi-static approximation all fields  are determined from  static Eqs.~(\ref{eq:Eq_for__psi}) and (\ref{eq:B_final}) by positions $r$ and momenta $p_r$ of macroparticles at given `time' $\xi$.
The source terms in these equations 
 do not have `time' derivatives of the fields or currents.


\section{Initial condition for distribution function}
 In this section we derive  the initial kick from the driver experienced by  plasma electrons. 

The ultrarelativistic driver~(\ref{eq:driver})  does not change the density $n_*$ because $n_{dr}-j_{dr}=0$. The driver is included in quasi-static equations by replacing $j_z\rightarrow j_z+j_{dr}$ in Eq.~(\ref{eq:S_r_}). 
Since $j_{dr}\propto \delta(\xi)$,
the $r$-coordinates of macroparticles  and their density   do not change during short interaction with the driver fields~\cite{barov_energy_2004}.   Therefore, $n_*|_{\xi=0+}=n_*|_{\xi=0-}=1$ and $\psi|_{\xi=0+}=0$.  From Eqs.~(\ref{eq:B_final}) and (\ref{eq:S_r_}), we conclude that 
$B_{\theta}\propto \delta(\xi)$. 
 It follows from Eqs.~(\ref{eq:EoM_1r})  that  momentum $p_r$  changes by a finite quantity during short `time'  $\xi-0$ to $\xi+0$.  Omitting  finite terms in the source~(\ref{eq:S_r_}),    we transform Eq.~(\ref{eq:B_final}) in the vicinity of the point $\xi=0$  to the Helmholtz equation with constant coefficients:
\begin{eqnarray}
\frac{\partial}{\partial r}\frac{1}{r} \frac{\partial}{\partial r}rB_{\theta}=B_{\theta}-\frac{\partial}{\partial r}j_{dr}. 
\label{eq:B_r_t_}
\end{eqnarray}
The  solution of this equation
$
B_{\theta}=-2QK_1(r)\delta(\xi) 
$, where $K_1(r)$ is the modified Bessel function of the second kind,
 determines the  initial transverse momentum  as a function of $r$:  
\begin{eqnarray} 
p_r|_{\xi=0+}=2Q K_1(r).
\label{eq:init_pr_}
\end{eqnarray}
 The corresponding   initial distribution function is: $f_*|_{\xi=0+}=\delta[p_r-2QK_1(r)]$. Note that  the initial `time' derivative of $\psi$ does not vanish $E_z|_{\xi=0+}=2QK_0(r)$, although $\psi|_{\xi=0+}=0$.    
	The initial longitudinal momentum and relativistic factor  are
\begin{eqnarray}
&&p_z|_{\xi=0+}=\gamma|_{\xi=0+}-1=\frac{1}{2}[2Q K_1(r)]^2.\label{eq:init_pz}
\end{eqnarray}
The initial  density of plasma electrons $n_e|_{\xi=0+}=\gamma|_{\xi=0+}=1+\frac{1}{2}[2Q K_1(r)]^2$ is apparently singular at $r=0$, but it turns out that   this  singularity does not affect  the characteristics of the plasma flow at $\xi>0$.

\section{Ballistic regime}
Ballistic regime takes place during  the initial stage of the plasma bubble formation ($\xi\ll 1$), when  electromagnetic fields generated by  plasma charges and currents do not yet affect the motion of the plasma electrons. These  electrons move with constant velocity ($dp_r/d\xi=0$ and $dr/d\xi=p_r$) vacating the area close to axis of the driver propagation.
 The ballistic trajectory $r=r(\xi)$  
 starting from the initial position $r_0$ is a straight line in the plane $(\xi,r)$  determined by
\begin{eqnarray} 
r=r_0+2QK_1(r_0)\xi.
\label{eq:bal_traj}
\end{eqnarray}
The condition $dr_0/dr=0$ determines the  envelope of these trajectories by the following parametric equations:
\begin{eqnarray}
&& r_{bal}=r_0 -K_1(r_0)/K_1'(r_0),
\label{eq:bal_traj2}\\
&& \xi=-1/[2QK_1'(r_0)], \label{eq:bal_traj1}
\end{eqnarray}
where  $r_0$ plays a role of  an independent parameter.   Note that  envelopes at different $Q$ can be obtained from each other by rescaling the axis $\xi$. At small $\xi$, the envelope radius scales as $r\propto \xi^{1/2}$ while at large $\xi$ the  radius changes logarithmically  $r\sim \ln \xi$.

This envelope creates a boundary of the fully evacuated area: 
$0<\xi$, $r<r_{bal}(\xi)$.
%
One could expect that electromagnetic fields excited during plasma electron motion should obstruct this motion, rendering  plasma bubbles of smaller radii: $r_b(\xi)<r_{bal}(\xi)$. However, we find  that   this is not  the case at large $Q$.

\section{PIC simulations of the electron plasma flow}

Based on  quasi-static approximation, we have developed in-house particle-in-cell (PIC) code  that is  similar to an electrostatic PIC code~\cite{Birdsall_book}: We define a grid that divides the  radial simulation domain $(0, R_{\max})$, where $R_{\max}\sim 100-300$,  into a set of small cells with size $\Delta r\sim 0.02$. 

Assuming that positions and momenta of macroparticles are known at the `moment' $\xi$, we calculate the the macroparticle density in each cell, solve Eq.~(\ref{eq:Eq_for__psi}), and determine $\psi$. 
After that we calculate the average `velocity' in each cell $\langle V\rangle=\langle p_r\rangle/(1+\psi)$ and then find $E_z$.
After calculating  $\langle V^2\rangle=\langle p_r^2\rangle/(1+\psi)^2$ and   $S$, we solve Eq.~(\ref{eq:B_final}) and find $B_{\theta}$. To avoid accumulation of round-off errors, we solve the Helmholtz equation for magnetic field by the method of Gaussian elimination   with partial pivoting~\cite{burden_numerical_2015}. After finding fields $\psi$, $\partial\psi/\partial r$, and $B_{\theta}$, we use Eqs.~(\ref{eq:EoM_1r}) to push macroparticles  to the next `moment' $\xi+\Delta \xi$, where  $\Delta\xi$ is the `time' step that  can be chosen by physical requirements at each moment $\xi$.   We use the adaptive Runge-Kutta method  for the `time' integration, and vary the 'time' step $\Delta\xi$   to accurately reproduce the frontal and rear parts of the plasma bubble where the bubble boundaries  are steep.

\subsection{Comparison of plasma bubbles at small and large driver charges}
\begin{figure}[t]
\centering
  \includegraphics[width=1.\columnwidth]{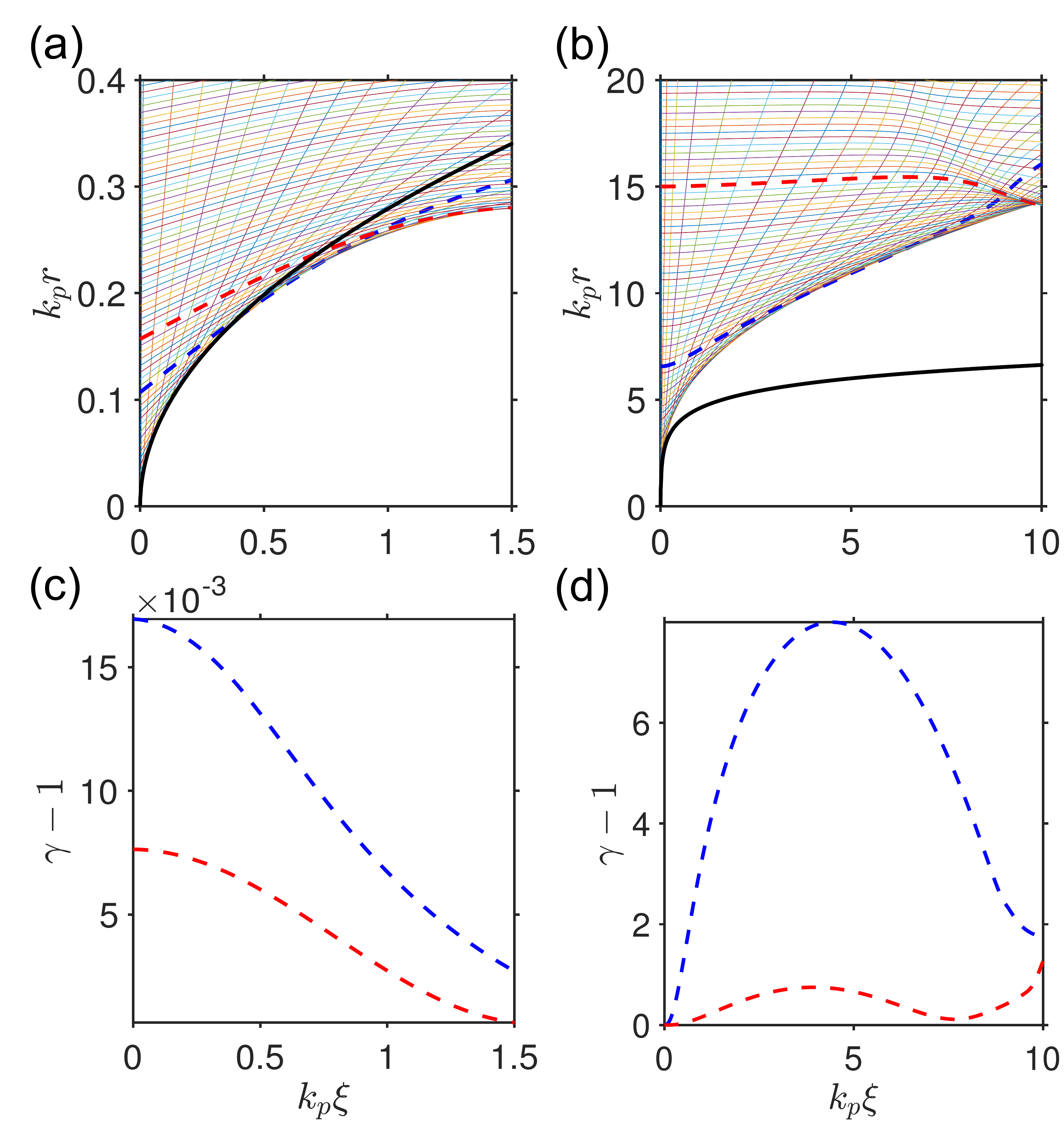}\\
\caption{Trajectories of electrons near the frontal part of the plasma bubble obtained from quasistatic PIC simulations: (a)  $Q=0.01$ and (b) $Q=25$. Blue and red dashed curves denote representative trajectories  touching the bubble boundary at different points. Panels (c)  Q=0.01 and (d) Q=25 show  change of the particle energy along these trajectories.
Black curves  are the envelopes of ballistic trajectories  determined by Eqs.~(\ref{eq:bal_traj2}) and (\ref{eq:bal_traj1}).  The driver is located at the point $(\xi,r)=(0,0)$.
 }
\label{fig:traj}
\end{figure}  

It is instructive to compare  the plasma bubbles at small and large values of $Q$.  Figure~\ref{fig:traj} shows trajectories of particles near the frontal part of the  bubble 
 at  $Q=0.01$ and $Q=25$.
It  also shows the envelopes of the ballistic trajectories.
 
When the driver charge is small, the electron velocities  in the vicinity of the bubble are also small ($v_r\sim 2Q/r\ll 1$ and $v_z\approx v_r^2/2\ll v_r$). The magnetic field can therefore be neglected, so that the electrons experience only attraction to the axis by  the bubble ions, a force proportional to $\partial_r \psi$. One can find then the transverse bubble radius $r_{\max}$ from simple  estimates. During particle motion, its initial kinetic energy   $T\sim  (2Q/r)^2/2$ is transformed to the potential energy of this particle $U\sim\frac{1}{4} r^2$ at the point  where the radius of the bubble reaches maximum value $r_{\max}$, see dashed  trajectories in Fig.~\ref{fig:traj} (a) and  corresponding  lines in Fig.~\ref{fig:traj} (c). 
Substituting $r\sim r_{\max}$ in the expression for $T$ and $U$ we find that $r_{\max}$ scales with  the charge as $r_{\max}\propto Q^{1/2}$. Note that the ion space charge puts the boundary of  the plasma bubble   below the 
 envelope of ballistic trajectories [black curve in Fig.~\ref{fig:traj} (a)]. 

A large charge driver initiates a relativistic flow of electron plasma and creates a bubble of large dimensions, see   Fig.~\ref{fig:traj} (b). 
If one were to estimate the bubble radius   by equating the initial kinetic energy of a particle to its potential energy at the bubble center, they would find  the bubble dimension close to the envelope of ballistic trajectories. In reality, the bubble boundary  lies considerably above this envelope 
despite the fact that the initial kinetic energy of the particles  between the dashed blue one  and the dashed red one  in Fig.~\ref{fig:traj} (b) is  small (due to exponential decrease of Bessel function in Eq.~(\ref{eq:init_pr_})) and can be neglected in all estimates. Moreover, instead of losing, these particles gain some kinetic energy during initial stage of bubble formation, see Fig.~\ref{fig:traj} (d). 

\begin{figure}[t]
\centering
	\includegraphics[width=0.95\columnwidth]{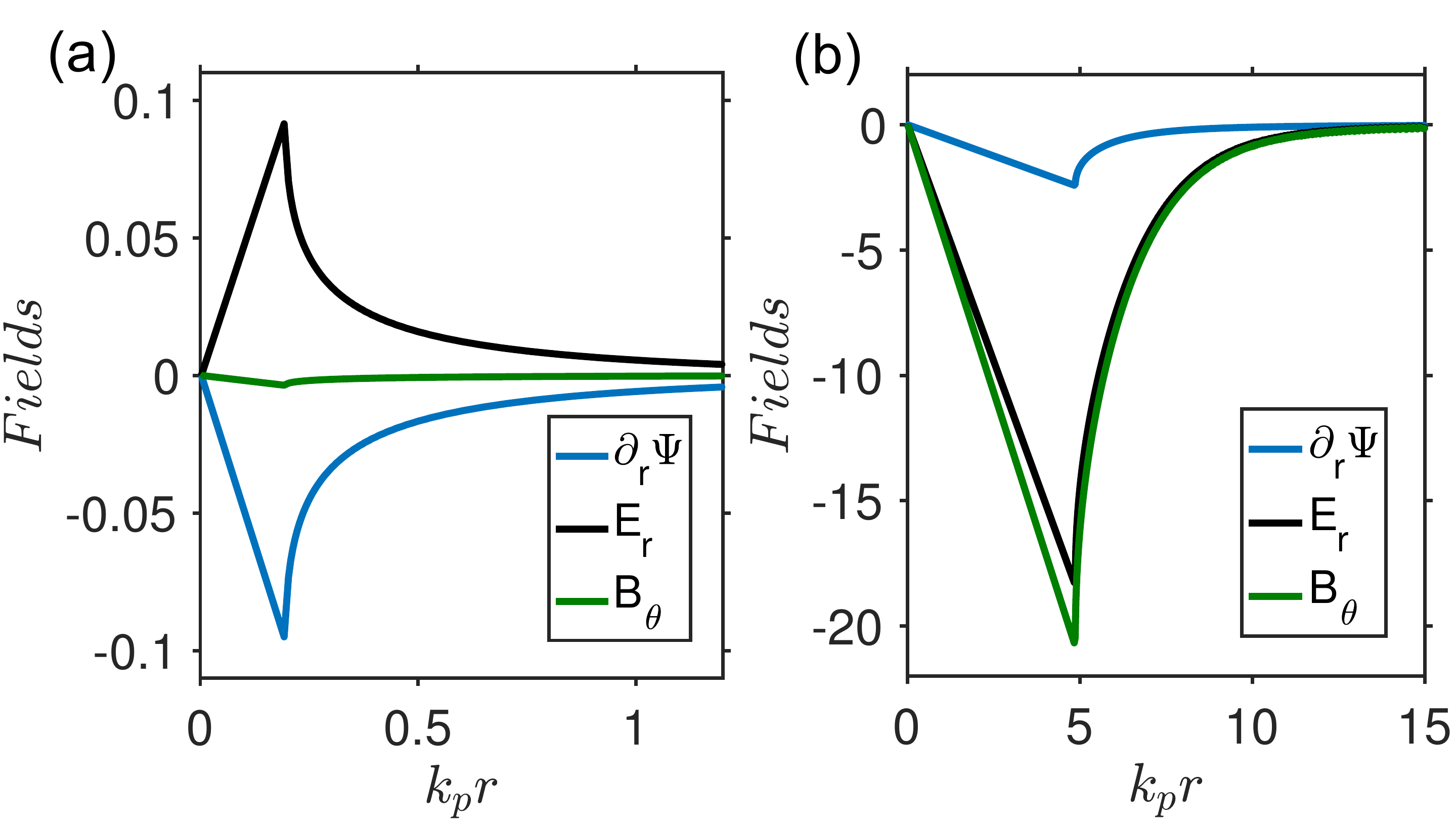}
	\caption{The radial electric field $E_r$, the radial gradient of the wakefield potential $\partial \psi/\partial r$ and azimuthal magnetic field $B_{\theta}$ as a function on radius $r$ at $\xi=0.5$: (a)  $Q=0.01$ and (b)  $Q=25$. Bubble boundaries are at $r\approx 0.19$ in (a) and $r\approx 5$ in (b). }
\label{fig:Electric_field}
\end{figure}

We can qualitatively explain  this effect  as follows.  The electric field in the bubble at small $Q$   
 is created only by the electron and ion charges $E_r\approx -\partial \psi/\partial r$. Due to the surplus of ions this field always attracts electrons to the axis ($-E_r<0$), see Fig.~\ref{fig:Electric_field} (a).
In contrast, the electric field  for large $Q$ has  a contribution from  the strong magnetic field $E_r=-\partial \psi/\partial r+B_{\theta}$, which determines the sign of the radial component ($-E_r>0$) at small $\xi$, see Fig.~\ref{fig:Electric_field} (b).    The  resulting motion of electrons is governed  by the radial Lorentz force $F_r=-E_r+v_zB_{\theta}$. 
Electrons with small energies (and small $v_z$) are repelled from the bubble axis ($F_r\approx -B_{\theta}>0$), while the energetic electrons ($v_z\approx 1$) are attracted, $F_r\approx \partial \psi/\partial r<0$. As a result,  energetic electrons lose their energy in the region near the bubble front  transferring part of it to  cold electrons.




\subsection{Plasma bubble and secondary plasma waves. }

\begin{figure}[t]
\centering
	\includegraphics[width=1\columnwidth]{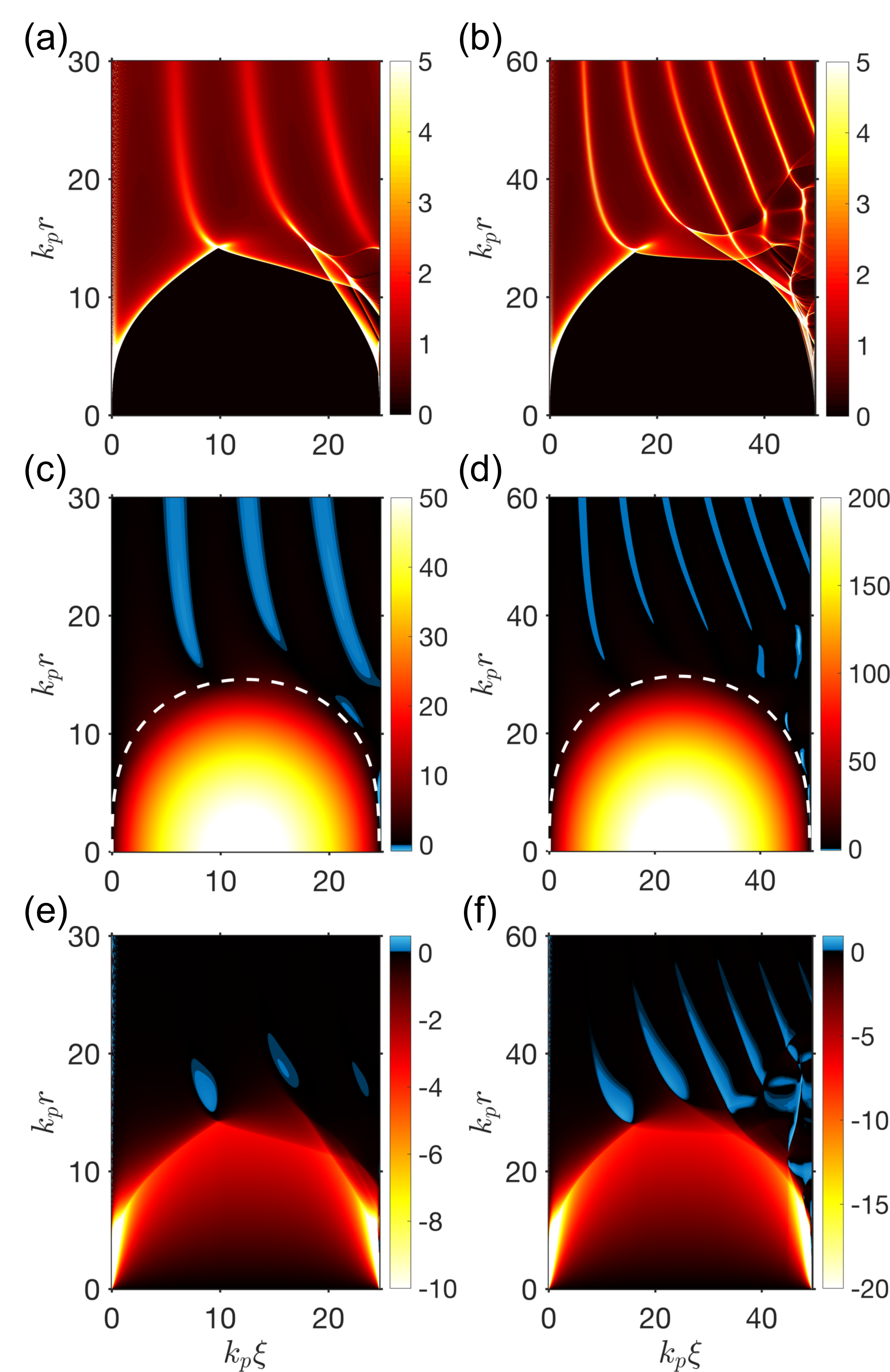}\\
	 \caption{Plasma characterisics at large driver charge $Q=25$ (frist column) and $Q=100$ (second column): density $n_e$ (upper panels), wakefield potential $\psi$ ( middle panels), and magnetic field $B_{\theta}$ (lower panels). Red color on  panels (e) and (f) denotes  negative value of the magnetic field.}
\label{fig:Q}
\end{figure}

Figure~\ref{fig:Q} shows  characteristics of the plasma flow 
 in the wake  of pointlike bunch of ultrarelativistic electrons with  $Q=25$ and $Q=100$.
The scales of the panels  
suggest that the bubble dimensions are proportional to $Q^{1/2}$. The  radius $R_b$ of the plasma bubble is slightly larger than half of the bubble length $L_b$;  the bubble resembles a deformed sphere. 
Note that 
the bubble boundary 
 has many kinks in its rear part where plasma streams inject electrons into the bubble (see panels (a) and (b) in Fig.~\ref{fig:Q}).  



The wakefield potential is positive inside  the bubble and its maximum value is $\psi\propto R_b^2\propto Q$. 
The dashed white lines in Fig.~\ref{fig:Q} (c) and (d) border the  areas where  $\psi$ is large. 
The magnetic field is negative inside  the  bubble (see 
 panels (e) and (f) in Fig.~\ref{fig:Q}). 
    It grows with the driver charge as $Q^{1/2}$.
 

\begin{figure}[t!!!]
\centering
	\includegraphics[width=1\columnwidth]
	{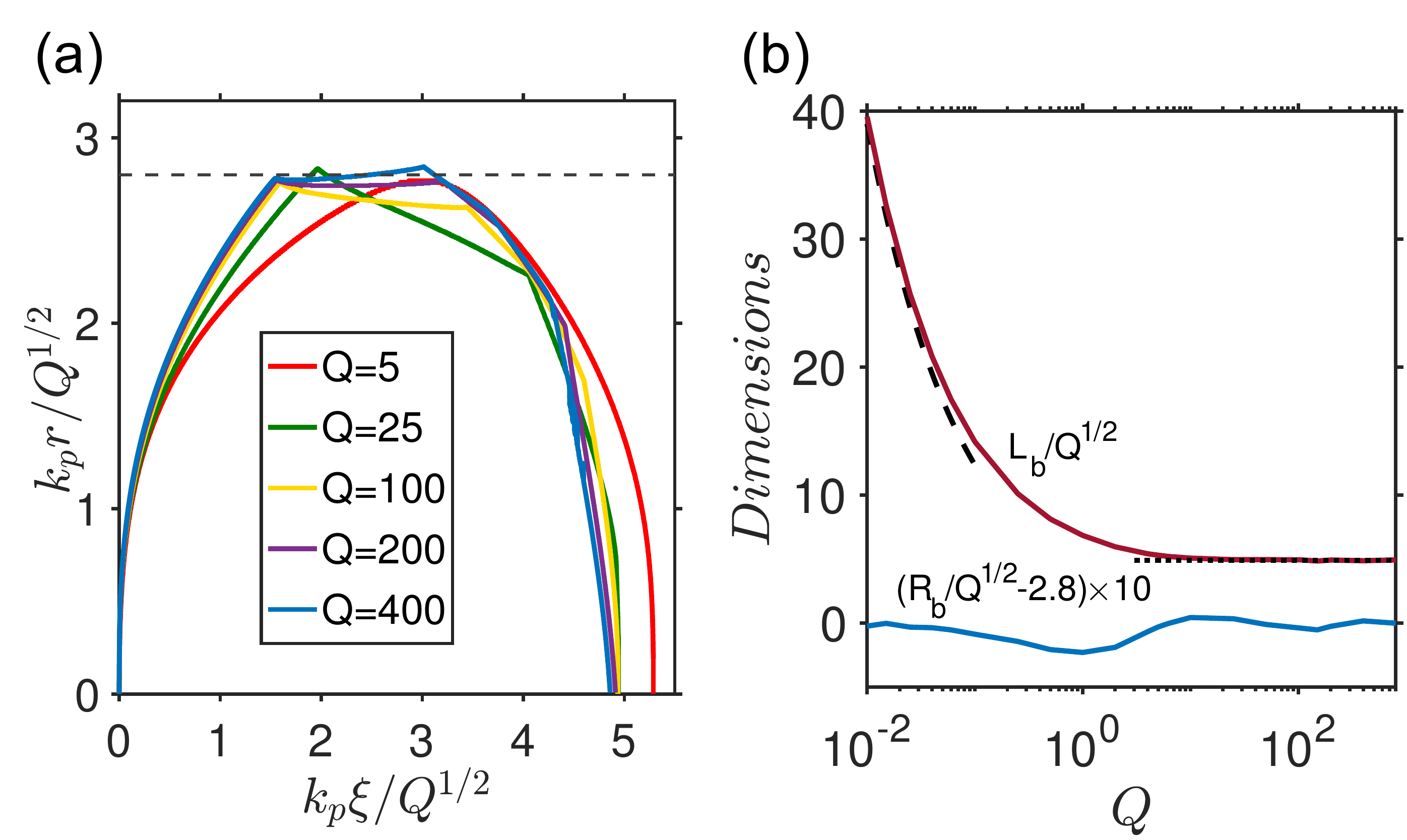}
\caption{(a) Boundaries of plasma bubble at large $Q$ , and (b) dependence of the radius and   length of the plasma bubble on $Q$  in the range $0.01\leq Q\leq 100$. Dashed line at the left corresponds to asymptote  $L_b=3.8$  at small $Q$, while the right dashed line to asymptote $L_b=4.9 Q^{1/2}$ at large $Q$.
 }
\label{fig:envelopes}
\end{figure}

 To verify the scaling laws, we have performed  PIC simulations for a range of  driver charges. Figure~\ref{fig:envelopes}~(a) shows plasma bubble boundaries at large $Q$. The  radius  and  length of the  bubble  are approximated at large $Q$ as $R_b \approx 2.8Q^{1/2}$ and $L_b\approx 4.9Q^{1/2}$. 

It is noteworthy that the ratio $R_b/Q^{1/2}$ is nearly constant within the explored range of the driver charges as indicated by  the  blue curve in Fig.~\ref{fig:envelopes}~(b). This ratio is 2.8 at small $Q$,  reaches a minimum of $2.5$ at $Q\approx 1$   and then returns to the value $2.8$ at large $Q$. The length of the plasma bubble has a  constant value~\cite{stupakov_wake_2016} at small $Q$ and it is approximated as $L_b\approx 4.9Q^{1/2}$ at large $Q$. One can see in Figure~\ref{fig:envelopes}~(a) exhibits some deviation of the bubble length from the $Q^{1/2}$-scaling already at $Q=5$.

Figure~\ref{fig:Q} also shows reveals
 oscillations of the plasma density  $\delta n_e$  outside the bubble;  these oscillations  are correlated with the narrow 
stream of  energetic particles expelled by the driver charge. Since the density of energetic (hot) electrons $n_h$ is small at large distances ($r\gg Q^{1/2}\gg 1$),    the oscillations of cold plasma are weak  and can be described  in the linear approximation ($\psi<<1$, $\gamma\approx 1$, $\delta n_e=\delta n_*-\psi\ll 1$).
Linearizing    Eqs.~(\ref{eq:EoM_1r}), (\ref{eq:Eq_for__psi}) and (\ref{eq:B_final}), we obtain:
\begin{eqnarray}
&&\frac{\partial^2}{\partial \xi^2}\delta n_e+\delta n_e=-n_h. 
\label{eq_lin1:n_e}
\end{eqnarray}

The stream of energetic electrons is only weakly perturbed by  the fields near the bubble. Consequently,  we can use  ballistic approximation to find  $n_h$. Omitting the intermediate  steps, we present the final approximate formula: 
\begin{eqnarray}
 n_h=\frac{2Q^2}{r^2}F\bigg(\frac{Q\xi}{r}\bigg),\quad F\equiv\frac{1}{1+5(Q\xi/r)^{3/2}}. 
\label{eq:app_n_h1}
\end{eqnarray}
Integrating Eq.~(\ref{eq_lin1:n_e}) with initial conditions 
$\delta n_e|_{\xi=0+}=0$ and  $(\partial\delta n_e/\partial \xi)|_{\xi=0+}=0$, we find  that, at large $r$,
its solution replicates  the  phase and amplitude of the plasma waves   observed in the simulations. As predicted by Eq.~(\ref{eq_lin1:n_e}), the plasma waves form  vertical striations with a period  $\Delta\xi=2 \pi$.  The number of striations     scales with the driver charge as $L_{st}\propto L_b/2\pi\propto Q^{1/2}$. Therefore the flow pattern in the striation area  lacks self-similarity.

  At small distances from the bubble, the waves become nonlinear,  and eventually brake. Merging with bubble boundary, the striations  create kinks that destroy boundary smoothness.  

One can show that small oscillations of the wakefield potential can be estimated as  $\psi\approx -\delta n_e$, see Fig.~\ref{fig:Q} (c) and (d), while the magnetic field does not oscillate at all in the linear approximation [see Fig.~\ref{fig:Q} (e) and (f)].


\section{Phenomenological models of plasma bubble at large $Q$}
As already  mentioned,   the plasma bubble in the wake of the pointlike electron driver  looks like a slightly deformed sphere when  $Q$ is large. Yet, this small deformation brings new properties that are very different from those of spherical bubble~\cite{Kostyukov_phenomenological_2004}. 
To demonstrate this difference, we develop  a simple model that captures the essential features of  the  bubble   observed in PIC simulations. 

\begin{figure}[t]
\centering 
	\includegraphics[width=0.85\columnwidth]
	{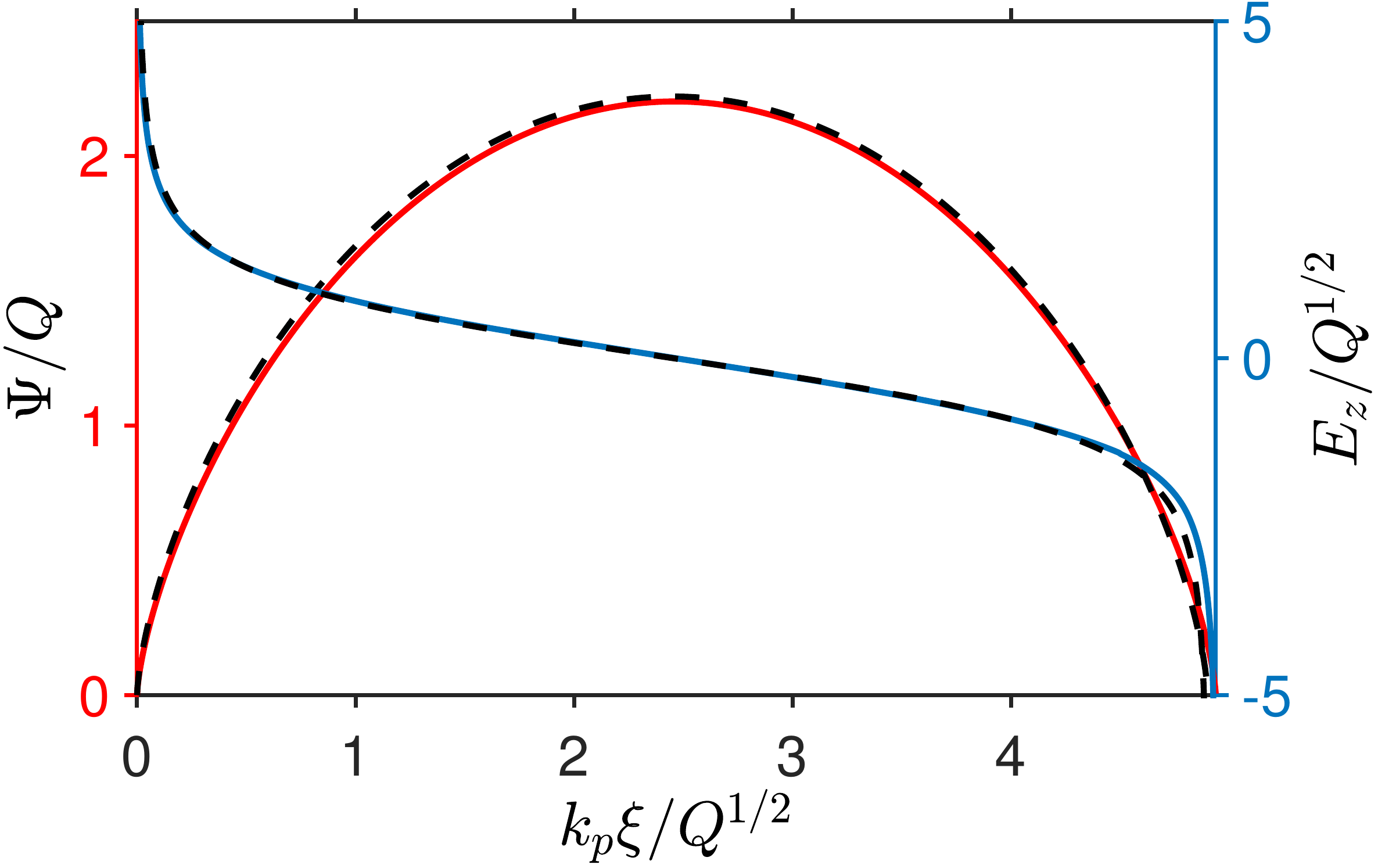}
   \caption{The wakefield potential $\psi$ 
	(red) and the longitudinal electric field $E_z$ 
	(blue)  on the bubble axis obtained from PIC simulations at $Q=100$. Dashed lines correspond to fitting curves $\bar{\psi}(\xi)=\bar{r}_b^2(\xi)/4$ and $d\bar{\psi}(\xi)/d \xi$, where $\bar{r}_b(\xi)$ is defined by Eq.~(\ref{eq:psi_boundary1}). }
	\label{fig:psi_boundary}
\end{figure}

The simulations show that electromagnetic energy is mostly stored inside  the bubble, 
where the electron density vanishes. To simplify the sheath structure, we assume that all fields vanish outside  the bubble. Integration of Eq.~(\ref{eq:Eq_for__psi}) with $n_*=0$ 
gives the wakefield potential inside  the bubble
\begin{eqnarray}
&&\psi(\xi,r)=\psi(\xi,0)-\frac{1}{4}r^2, 
\label{eq:psi_in_bubble}
\end{eqnarray}
 where $\psi(\xi,0)$ is an on-axis potential. It turns out that our  simulation results suggest a very accurate approximation for the wakefield potential on the bubble axis.  More specifically, 
%
$\psi(\xi,0)\cong \bar{\psi}(\xi)\equiv \frac{1}{4}\bar{r}_b(\xi)^2$ with
$\bar{r}_b(\xi)$  defined by  
 one-parameter equation 
\begin{eqnarray}
\frac{\bar{r}_b^3}{R_b^3}+{\bigg(1-\frac{\xi}{\xi_c}\bigg)^2}=1
\label{eq:psi_boundary1}
\end{eqnarray}
where $\xi_c\equiv\frac{1}{2}L_b\approx (6Q)^{1/2}$ 
and $R_b\approx 3Q^{1/2}-0.2$. As  seen in
 Fig.~\ref{fig:psi_boundary}, Eq.~(\ref{eq:psi_boundary1})  gives also a good approximation 
for $E_z(\xi,0)$. 
Function $\bar{r}_b(\xi)$ can be interpreted as a radius at which the wakefield potential $\psi(\xi,r)$ given by Eq.~(\ref{eq:psi_in_bubble}) formally vanishes. Such $\psi$-boundaries of the plasma bubble are smooth and denoted in Fig.~\ref{fig:Q} (c) and (d) by white dashed lines. 

All fields inside  the    bubble can now be expressed  through  $\bar{r}_b(\xi)$:
\begin{eqnarray}
&&\psi(\xi,r)
= \bar{\psi}(\xi)-\frac{1}{4}r^2, 
\label{eq:psi_in_bubble11}\\
&&\frac{\partial \psi}
{\partial r}=-\frac{1}{2}r, \quad
E_z=\frac{d \bar{\psi}(\xi)}{\partial \xi},\quad
B_{\theta}=\frac{1}{2}\frac{\partial E_z}{\partial \xi}r,
\label{eq:Bth_in_bubble}
\end{eqnarray}
where $r\leq \bar{r}_b(\xi)$, $0<\xi<2\xi_c$, and $\bar{\psi}(\xi)=\frac{1}{4}R_b^2[1-(1-\xi/\xi_c)^2]^{2/3}$.
Outside  the bubble, these fields  are set to  zero.
We   see from these expressions, once again,  that $\psi$ is proportional to $Q$ and $\partial_r\psi$,  $E_z$, and $B_{\theta}$ are proportional to $Q^{1/2}$. These scalings do not hold in  striations 
but the fields  are relatively weak outside  the bubble.

A  spherical model of the bubble  developed 
for a laser driver 
in Ref.~\cite{Kostyukov_phenomenological_2004} suggests
\begin{eqnarray}
&&\frac{\bar{r}_b^2}{R_b^2}+{\bigg(1-\frac{\xi}{R_b}\bigg)^2}=1, \label{eq:standard_boundary}
\end{eqnarray}
where $R_b$ is  the bubble radius and $0<\xi<2R_b$.  In this case  $\psi(\xi,r)= \frac{1}{4}[R_b^2-(R_b-\xi)^2-r^2]$ inside  the bubble and  $\psi(\xi,r)= 0$ outside.

Equations~(\ref{eq:psi_boundary1}) and (\ref{eq:standard_boundary}) look very similar at first sight, especially if one takes into account that $\xi_c$ only slightly smaller than $R_b$. Despite this similarity, the fields are dramatically different at the bubble frontal parts. In the large charge model~(\ref{eq:psi_boundary1}), we find that    $\bar{r}_b\approx R_b(2\xi/\xi_c)^{1/3}$ at $\xi\ll R_b$, 
and the magnetic field
is very large: 
$B_{\theta}\approx -\frac{1}{9}r(R_b^6/\bar{r}_b^4\xi_c^2)$. 
This field determines the radial electric field
\begin{eqnarray}
E_r=-\partial \psi/\partial r+B_{\theta}\approx B_{\theta}
\label{eq:E_r_in_bubble}
\end{eqnarray}
More accurate estimate shows that the radial electric field is negative up to   $\xi=0.3\xi_c$.  In contrast,  $B_{\theta}\approx -\frac{1}{4}r$ in the spherical model, and the radial field is always positive $E_r=\frac{1}{4}r$. 
\begin{figure}[t!]
\centering
	\includegraphics[width=1\columnwidth]{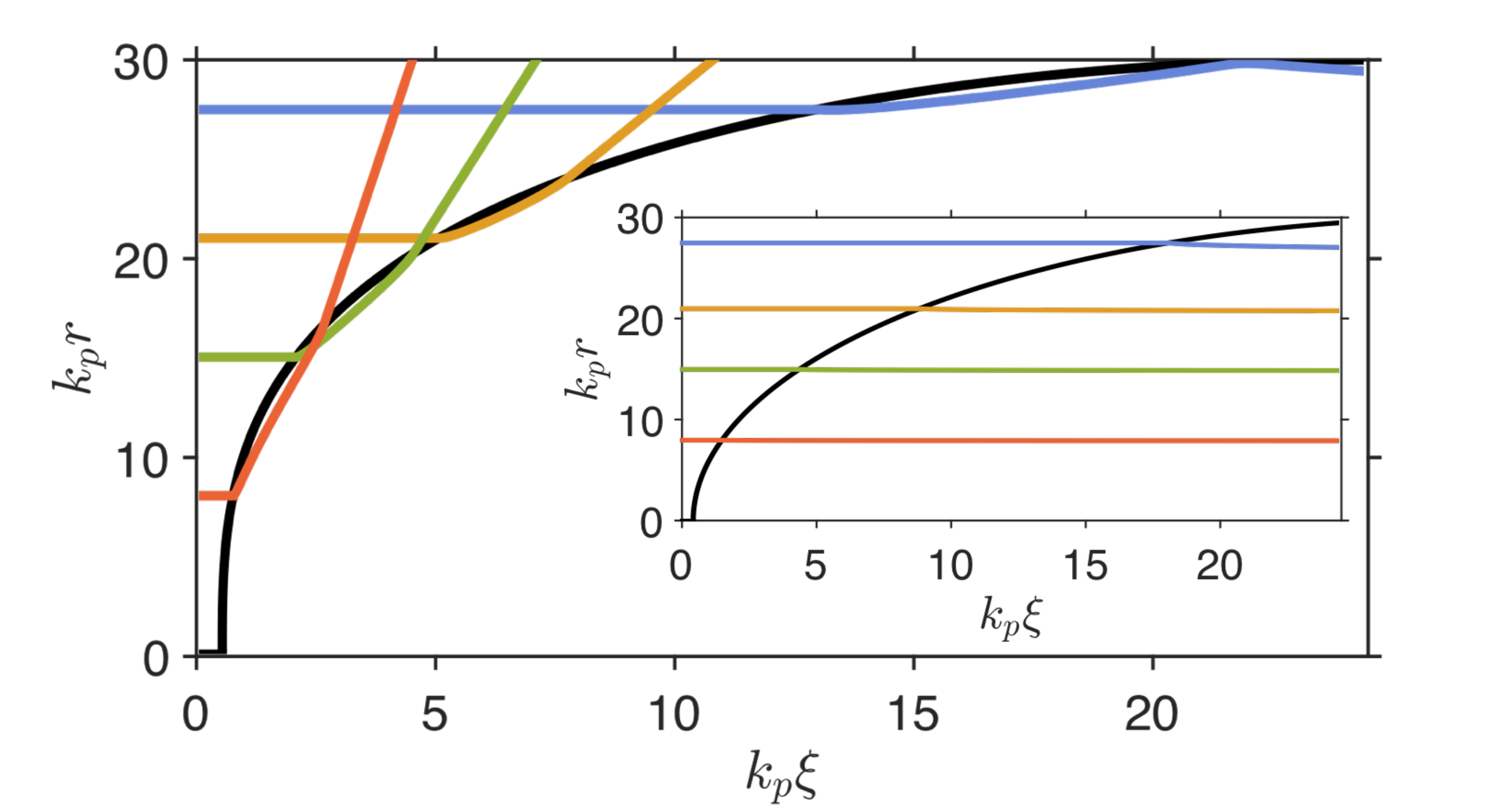}
   \caption{Trajectories of of test electrons initially placed at rest at different distances from the bubble axis.  Electrons are  repelled by the bubble in the large charge model~(\ref{eq:psi_boundary1}) while   the same electrons  penetrate into the spherical bubble~(\ref{eq:standard_boundary}) (inset). In both models $R_b\approx 30$.}
	\label{fig:test_electrons}
\end{figure}
As a result,   cold electrons experience strong repulsion  at the frontal part of the bubble  in the large charge model,  
while the same electrons are pulled inward in  the spherical bubble case (see Fig.~\ref{fig:test_electrons}).


One can get a better understanding of the structural  difference between these bubbles by examining a part of   energy~(\ref{eq:conserv_energy1})  associated with the wakefield potential: 
\begin{eqnarray}
U_{pot}=2\pi \int_{0}^{\infty}rdr\frac{[\nabla\psi(\xi,r)]^2}{2}
\label{eq:U_tot}
\end{eqnarray}
where  $(\nabla\psi)^2=E_z^2+(\partial\psi/\partial r)^2$.  At  $\xi<<R_b$, the contribution from the radial gradient of the wakefield potential can be neglected. Hence, $(\nabla\psi)^2\approx E_z^2$ and  $U_{pot}\approx \frac{1}{2}\pi E_z^2 r_b^2$.
Since $E_z^2\approx R_b^2/4$  and $r_b^2\approx 2R_b\xi$, the initial potential energy in the spherical model~(\ref{eq:standard_boundary}) vanishes: $U_{pot}|_{\xi=+0}=0$.  In contrast,   $E_z^2\propto 1/r_b^2$ and   
the initial potential energy   is finite in the large charge model~(\ref{eq:psi_boundary1}): $U_{pot}\approx 21 Q^2$. This potential energy 
determines subsequent evolution of the plasma bubble: change in $E_z$ creates a displacement current $\partial E_z/\partial \xi$ that generates strong magnetic field $B_{\theta}$ which, in turn, induces the radial electric field $E_r$~(\ref{eq:E_r_in_bubble}). 



\begin{figure}[t]
\centering
	\includegraphics[width=1\columnwidth]{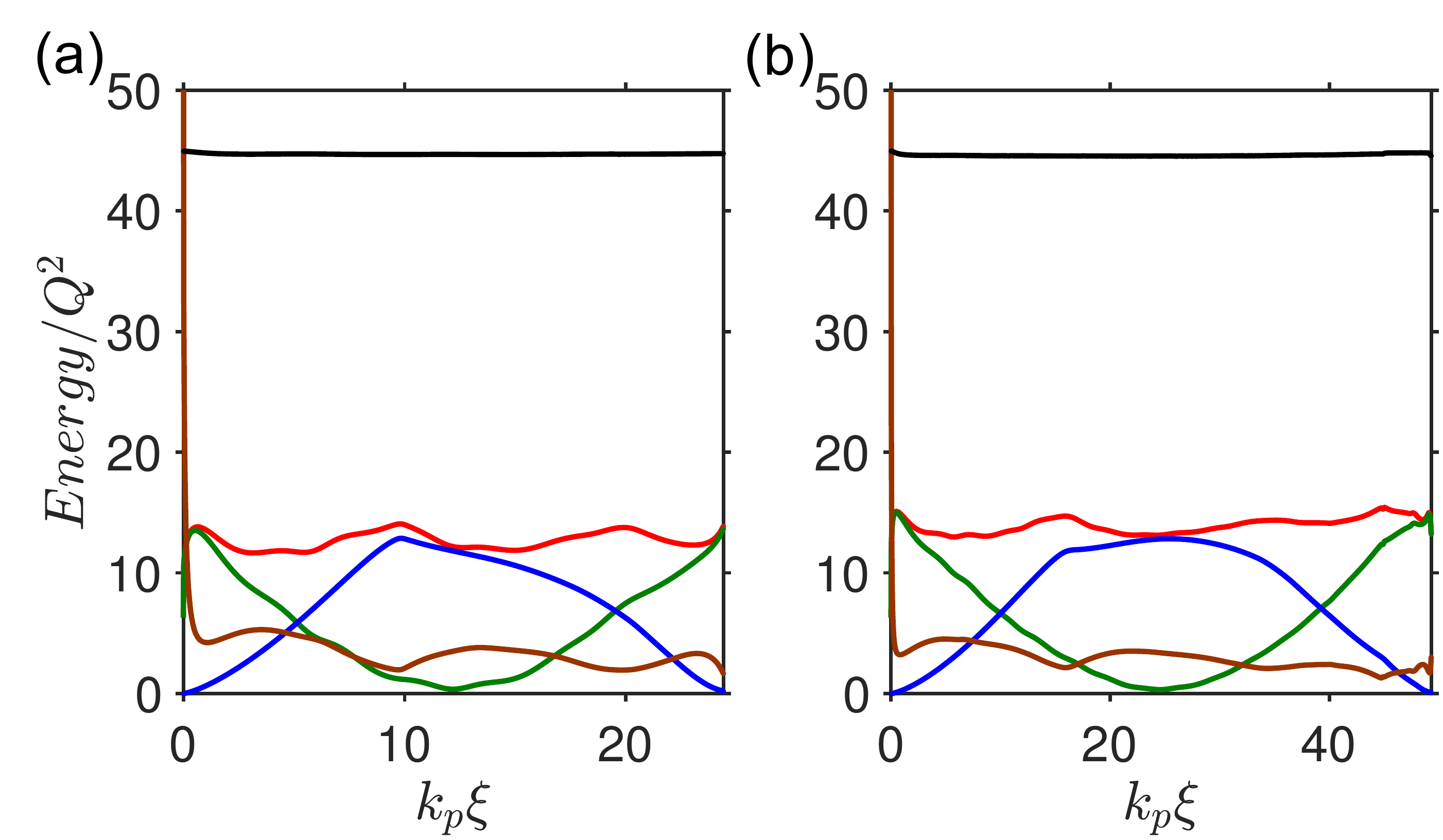}\\
	 \caption{Change of the energy with `time' $\xi$  for (a) $Q=25$ and (b) $Q=100$ observed in PIC simulations:  the kinetic energy of macroparticles in the region $r<2R_b$(brown), the potential energy stored in $E_z$ (green), potential energy stored in $\partial_r\psi$ (blue),  total potential energy (red), and total energy of the system normalized by $2Q^2$ (black).}
\label{fig:energy_on_xi}
\end{figure}

Analysis of the bubble energy reveals additional interesting aspects of  the bubble behind the large charge driver.   Figure~\ref{fig:energy_on_xi} shows that  the total energy of the system  in the quasistatic PIC  simulations (plotted by black line) remains constant in agreement with Eq.~(\ref{eq:conserv_energy1}). 
%
The potential energy  in the $E_z$-component of the electric field approximately doubles during very short `time' $\xi<<1$ via energy transfer from  macroparticles. After that, $E_z$ field energy slowly, while the energy attributable to  the radial gradient of the wakefield potential increases  till $\xi<\xi_c$. 
Unexpectedly,   variation of 
the total potential energy (plotted by red line) with `time' $\xi$ 
is  relatively small (within $17\%$). 
Note that the fields are nearly  zero outside the bubble and   $E_z=\frac{1}{2}\bar{r}_b (d\bar{r}_b/d\xi)$ and $\partial \psi/\partial r=-\frac{1}{2}r$  inside the bubble.  Given that   variation of the potential energy  is relatively small, one can now write the energy balance equation as
 \begin{eqnarray}
\frac{1}{8}\pi \bar{r}_b^4 \Big(\frac{d\bar{r}_b}{d\xi}\Big)^2+\frac{1}{16}\pi\bar{r}_b^4  =\frac{1}{16}\pi R_b^4
\label{eq:En_const}
\end{eqnarray}
where $\frac{1}{16}\pi R_b^e$ is the potential energy at  $\xi=\xi_c$. 
A straighforward solution of this equation shows that the bubble  is a deformed sphere
with half length $\xi_c\approx 0.85 R_b$ that is slightly longer than the length obtained from PIC simulations.  Furthermore, differentiation of Eq.~(\ref{eq:En_const}) with respect  to $\xi$ gives 
\begin{eqnarray}
\bar{r}_b \frac{d^2\bar{r}_b}{d\xi^2}+2\Big(\frac{d\bar{r}_b}{d\xi}\Big)^2+1=0.
\label{eq:Lu}
\end{eqnarray}
The same equation follows from the
  model developed by Lu et al.~\cite{Lu_theory_2006}. Therefore, despite the fact that the  boundary of the plasma bubble created by the pointlike  driver is composed of many electron trajectories, the Lu model is supported by the approximate conservation  of the  potential energy.  

We conclude this section by noting that  integral~(\ref{eq:conserv_energy1}) logarithmically diverges at the lower limit at $\xi=0+$, because $\langle\gamma-1\rangle|_{\xi=0+}\approx 1/r^2$ at small $r$.  For this reason,  the total energy of the system increases with increase of simulation resolution. But this increase does not influence   the plasma electron flow and does not change the `time'-evolution of the potential energy 
at sufficiently small size of the radial cell: $\Delta r\ll 1$. 

\section{Summary}
We have advanced the nonlinear wakefield theory in  quasistatic approximation and developed a novel PIC code in which all fields are determined  from  static equations.
More specifically, the wakefield potential is found from the Poisson equation  and the magnetic field is found from the Helmholtz equation. The  source terms in these equations depend only on radial  positions and momenta of macroparticles.   This approach can be straightforwardly generalized to the case of  nonaxisymmetric plasma flows, see Appendix C. 
	
We have characterized the plasma flow in the wake of a  pointlike bunch of ultrarelativistic electrons. The radius and the length of the plasma bubble are found for varying values of the bunch charge. We show that the normalized bubble radius $R_b/Q^{1/2}$ is approximately equal to $2.8$ for small and large values of $Q$, and  only slightly deviates  from $2.8$ at $Q\sim 1$. The normalized length of the bubble has a constant value 3.8 for $Q\ll 1$, and grows    as $L_b\approx 4.9Q^{1/2}$ at  $Q\gg 1$. 
	


Our 
phenomenological models for large $Q$ explain how small deviation of the bubble from the spherical shape can dramatically change the bubble properties.
 by comparing a large charge model~(\ref{eq:psi_boundary1}) with spherical one~(\ref{eq:standard_boundary}). 
These  models reveal  the governing role
 of the potential energy~(\ref{eq:U_tot})  on the bubble  structure that reflects  transformation from the energy stored in the  field $E_z$ to the energy stored in the field $\partial_r\psi$, and vice versa.

Our simulations and analytical theory reveal plasma waves  excited by energetic particles at large distances from the bubble.
 The amplitude of these waves becomes large near the bubble  giving rise to  wavebreaking. When present at the bubble boundary, striations of plasma waves create kinks destroying boundary smoothness. 


This research was supported by DOE grants DE-SC0007889 and DE-SC0010622, and by an AFOSR grant FA9550-14-1-0045. B. Breizman was  supported by the U.S. Department of Energy Contract
No. DEFG02-04ER54742.

\begin{appendix}
\section{Vlasov's equation in the cylindrically symmetric form.}
 It is convenient to write Eq.~(2)  in the form: 
\begin{eqnarray}
\frac{\partial }{\partial t}{f_e} +\frac{\partial}{\partial {\bf{R}}}\cdot\bigg(\frac{\partial H}{\partial {\bf{P}}} f_e\bigg)-
\frac{\partial }{\partial {\bf{P}}}
\cdot\bigg(\frac{\partial H}{\partial {\bf{R}}}f_e\bigg)=0,
\label{eq:Vlasov1_1}
\end{eqnarray}
We replace $\partial  /\partial t\rightarrow \partial  /\partial \xi$ and $\partial  /\partial z\rightarrow -\partial  /\partial \xi$,  substitute the electron distribution function $f_e$ using Eq.~(6) and then integrate Eq.~(\ref{eq:Vlasov1_1}) over longitudinal and azimuthal momenta. Taking into account that  the function $\delta(H-P_z-1)$  integrated  over $P_z$  produces the factor $1/(1-v_z)=\gamma/(1+\psi)$, we obtain:
\begin{eqnarray}
\frac{\partial }{\partial \xi}{r f_*}+\frac{\partial}{\partial {r}}\bigg(\frac{\partial H}{\partial {p_r}}\frac{\gamma rf_*}{1+\psi}\bigg)-
\frac{\partial }{\partial {p_r}}
\bigg(\frac{\partial H}{\partial {r}}\frac{\gamma rf_*}{1+\psi}\bigg)=0,\,\,\,\,
\label{eq:Vlasov1_2}
\end{eqnarray}
Noting that Hamiltonian $H_*$ defined by Eq.~(8) satisfies the relations:
\begin{eqnarray}
\frac{\partial H_*}{\partial {p_r}}=\frac{\partial H}{\partial {p_r}}\frac{\gamma }{1+\psi},\quad
\frac{\partial H_*}{\partial {r}}=\frac{\partial H}{\partial {r}}\frac{\gamma }{1+\psi},
\label{eq:Vlasov1_3}
\end{eqnarray}
one can straightforwardly transform Eq.~(\ref{eq:Vlasov1_2}) to Eq.~(7).
\section{Conservation of energy of macroparticles.}
Energy conservation in differential form is given by
\begin{eqnarray}
\frac{\partial}{\partial t}\bigg[W+\frac{1}{2}(E^2+B^2)\bigg]+\nabla\cdot ({\bf{J}}+{\bf{E}}\times {\bf{B}})=0, 
\label{eq:energy_conser_A}
\end{eqnarray}
where  
$W=\int d^3{\bf{p}} f_e(\gamma-1)$ is the  density of electron kinetic energy
 and  ${\bf{J}}$  is its flux. Replacing $t$ and $-z$ by $\xi$
and performing integration in the transverse plane, we obtain
\begin{eqnarray}
\frac{\partial}{\partial\xi}\int d^2{\bf{r}}_{\perp}\bigg[W-J_z+\frac{E^2+B^2}{2}-[{\bf{E}}\times {\bf{B}}]_z\bigg]=0. 
\label{eq:continuity_energy}
\end{eqnarray}
One can directly  check that  $W-J_z=n_*\langle \gamma-1\rangle$ and $\frac{1}{2}(E^2+B^2)-[{\bf{E}}\times {\bf{B}}]_z=\frac{1}{2}[E_z^2+(\partial_r\psi)^2]$ in the cylindrically symmetric case, i.e., Eq.~(11) follows from Eq.~(\ref{eq:continuity_energy}). 
\section{Nonaxisymmetric plasma flows} 
In this section we present a condensed derivation of the basic equations under quasistatic approximation for nonaxisymmetric plasma flows. The current   density of the driving electron  beam   is given now by
\begin{eqnarray}
j_{dr}=
4\pi Q\delta(x)\delta(y)\delta(t-z).\label{eq:driver}
\end{eqnarray}
Assuming that the distribution function
$f_e$ and electromagnetic fields depend on t and z only
through a combination $\xi=t-z$, we find that $H-P_z-1=0$. The distribution function of electrons can be expressed in the form
\begin{eqnarray}
f_e(t,{\bf{R}},{\bf{P}})=f_{*}(\xi,{\bf{r}},{\bf{P}}_{\perp})\delta(H -P_z-1),
\label{eq:substitute_A}
\end{eqnarray}
where $f_*$ represents now a distribution function    of macroparticles performing two dimensional  motion in the $(x,y)$-plane and ${\bf{r}}\equiv(x,y)$.
Substituting Eq.~(\ref{eq:substitute_A}) into (\ref{eq:Vlasov_3D}), we find that $f_*$ satisfies the  Vlasov equation:
\begin{eqnarray}
&&\frac{\partial {f}_*}{\partial \xi} +\frac{\partial H_*}{\partial {\bf{P}}_{\perp}}\cdot\frac{\partial f_*}{\partial {\bf{r}}}-\frac{\partial H_*}{\partial {\bf{r}}}\cdot\frac{\partial f_*}{\partial {\bf{P}}_{\perp}}=0,\label{eq:Vlasov_2D_A}
\end{eqnarray}
where 
\begin{eqnarray}
 H_*=\frac{1+({\bf{P}}_{\perp}+{\bf{A}}_{\perp})^2+(1+\psi)^2}{2(1+\psi)}-\psi-A_z\label{eq:H_2D_A}
\end{eqnarray}
is the  Hamiltonian for the two-dimensional motion in the $(x,y)$-plane and $\psi=\phi-A_z$ is the wakefield potential.  The trajectory of an individual particle in the phase space $({\bf{r}},{\bf{P}}_{\perp})$  is determined by equations of motion
\begin{eqnarray}
&&\frac{d{\bf{P}}_{\perp}}{dt}=-\frac{\partial H}{\partial {\bf{r}}}, \quad \frac{d{\bf{r}}}{dt}=\frac{\partial H}{\partial {\bf{P}}_{\perp}}.
\label{eq:EoM_2D_A}
\end{eqnarray}
By replacing ${\bf{P}}_{\perp}={\bf{p}}_{\perp}-{\bf{A}}_{\perp}$, these equations  can be written in the form
\begin{eqnarray}
&&\frac{d}{d\xi}{\bf{p}}_{\perp}=\frac{\gamma\nabla_{\perp}\psi}{1+\psi} +[ {\bf{e}}_z\times {\bf{V}}]B_z+[{\bf{e}}_z\times {\bf{B}}_{\perp}],
\label{eq:EoM_111}\\
&&\frac{d}{d\xi}{\bf{r}}_{\perp}\equiv {\bf{V}}_{\perp}=\frac{1}{1+\psi}{\bf{p}}_{\perp} \label{eq:EoM_112_A}
\end{eqnarray}
where  $\gamma=[{1+{\bf{p}}_{\perp}^2+(1+\psi)^2}]/{2(1+\psi)}$ is the relativistic factor, and ${\bf{V}}\equiv {\bf{p}}_{\perp}/({1+\psi})$ is the particle `velocity' in $(x,y)$-plane. Integration of  Eq.~(\ref{eq:Vlasov_2D_A}) over  $P_x$ and $P_y$ gives the  continuity equation:
\begin{eqnarray}
&&\frac{\partial}{\partial \xi}n_{*}=
-\nabla_{\perp}\cdot (n_*\langle{\bf{V}}\rangle), 
\label{eq:continuity}
\end{eqnarray}
where  $n_{*}=\int dP_xdP_y f_*$ is the density of macroparticles and  the brackets denote averaging over transverse momentum $\langle {\bf{V}}\rangle={{n}}_{*}^{-1}\int dP_xdP_y f_{*}{\bf{V}}$.

We also note that  the density and the current density of plasma electrons can be expressed through the distribution function  $f_*$ as:
 \begin{eqnarray}
n_e=\frac{n_*\langle \gamma \rangle}{1+\psi},\quad {\bf{j}}_{\perp}=n_*\langle {\bf{V}}\rangle, 
\quad j_z=\frac{n_*\langle p_z\rangle}{1+\psi},\label{eq:Jr}
\end{eqnarray}
 where $\langle p_z\rangle=\langle \gamma \rangle-\psi-1=[1+\langle {\bf{p}}_{\perp}^2\rangle-(1+\psi)^2]/(2(1+\psi)$. Besides, $n_e-j_z=n_*$.

Using the similar transformations as in Section~II, we obtain the following equations  
\begin{eqnarray}
&&\Delta_{\perp} \psi=n_{*}-1, \label{eq:EqM_N8_AA}\\
&&\Delta_{\perp}E_z=-\nabla_{\perp}\cdot{\bf{j}}_{\perp},
\label{eq:EqM_N8_A}\\
&&\Delta_{\perp}B_z={\bf{e}}_z\cdot[\nabla_{\perp}\times {\bf{j}}_{\perp}], \label{eq:Mag_z_A}\\
&&\Delta_{\perp}{\bf{B}}_{\perp}=-[{\bf{e}}_z\times\nabla_{\perp}j_z]
-\bigg[{\bf{e}}_z\times\frac{\partial }{\partial \xi} {\bf{j}}_{\perp}\bigg]. \label{eq:Mag_x_A}
\end{eqnarray}
The solutions of Poisson Eqs.~(\ref{eq:EqM_N8_AA}) - (\ref{eq:Mag_z_A}) must vanish at $r\rightarrow\infty$.  To obtain a closed form of Eq.~(\ref{eq:Mag_x_A}), we  multiply the Vlasov Eq.~(\ref{eq:Vlasov_2D_A}) 
by the 'velocity' ${\bf{V}}_{\perp}=\partial H/\partial{\bf{P}}_{\perp}$ and integrate it over momentum. After straightforward calculations we find
\begin{eqnarray}
\frac{\partial}{\partial\xi}{\bf{j}}_{\perp}=n_{*}\langle{\bf{a}}\rangle-\frac{\partial}{\partial x}n_{*}\langle V_x{\bf{V}}_{\perp}\rangle-\frac{\partial}{\partial y}n_{*}\langle V_y{\bf{V}}_{\perp}\rangle.
\label{eq:FluxPres}
\end{eqnarray}
where ${\bf{a}}\equiv {d^2{\bf{r}}_{\perp}}/{d\xi^2}$ is the particle 'acceleration':
\begin{eqnarray}
&&{\bf{a}}=\frac{[{\bf{e}}_z\times {\bf{B}}_{\perp}]}{1+\psi}+\frac{[{\bf{e}}_z \times  {\bf{V}}_{\perp}]B_z}
{(1+\psi)}
+\tilde{\bf{a}},\\
&&\tilde{\bf{a}}=\frac{\gamma \nabla_{\perp}\psi}{(1+\psi)^2}-\frac{{\bf{V}}}{1+\psi}\bigg(E_z+{\bf{V}}\cdot
\nabla_{\perp} \psi\bigg).
\label{eq:accel_A}
\end{eqnarray}

Substituting ${\partial {\bf{j}}_{\perp}}/{\partial\xi}$ from Eq.~(\ref{eq:FluxPres}) to Eq.~(\ref{eq:Mag_x_A}) we obtain Helmholtz equation describing the transverse magnetic field:
\begin{eqnarray}
&&\Delta_{\perp}{\bf{B}}_{\perp}=\frac{n_{*}}{1+\psi}{\bf{B}}_{\perp}- [{\bf{e}}_z \times{\bf{S}}]
\label{eq:Main_Eq_A}
\end{eqnarray}
with the source  
\begin{eqnarray}
&&{\bf{S}}= \frac{[{\bf{e}}_z \times  n_*\langle{\bf{V}}\rangle]B_z}
{(1+\psi)}+n_{*}\langle\tilde{\bf{a}}\rangle-\nonumber\\
&&\frac{\partial}{\partial x}n_{*}\langle V_x{\bf{V}}_{\perp}\rangle-\frac{\partial}{\partial y}n_{*}\langle V_y{\bf{V}}_{\perp}\rangle +\nabla_{\perp}j_z.
\label{eq:FluxCurr1_A}
\end{eqnarray}
The solutions of Eq.~(\ref{eq:Main_Eq_A})  must vanish at $r\rightarrow\infty$.
Thus, in quasi-static approximation all fields are determined
from static Eqs.~(\ref{eq:EqM_N8_AA}) - (\ref{eq:Mag_z_A})  and (\ref{eq:Main_Eq_A}) by positions  and momenta
 of macroparticles in the plane $(x,y)$ at given `time' $\xi$. The source
terms in these equations do not have `time' derivatives of
the fields or currents.

When these equations are incorporated in the PIC code, it is convenient to replace an averaging over transverse components of the canonical momentum   by  the averaging over corresponding  components of the kinetic momentum. Indeed, since ${\bf{P}}_{\perp}={\bf{p}}_{\perp}-{\bf{A}}_{\perp}$ and averaging is performed at given point ${\bf{r}}$, we obtain 
\begin{eqnarray}
\frac{1}{n_*}\int dP_xdP_yf_*(\xi,{\bf{r}},{\bf{P}}_{\perp}){\bf{V}}= 
\nonumber\\
\frac{1}{n_*}\int dp_xdp_y
f_*(\xi,{\bf{r}},{\bf{p}}_{\perp}-
{\bf{A}}_{\perp}){\bf{V}}.
\label{eq:last_Eq_A}
\end{eqnarray}

\end{appendix}

  \nocite{*}

 

 \end{document}